\documentclass[%
 reprint,
superscriptaddress,
 amsmath,amssymb,
 aps,prl,
]{revtex4-2}

\usepackage[T1]{fontenc}
\usepackage[utf8]{inputenc}
\usepackage{graphicx}
\usepackage{amsmath}
\usepackage{xcolor}
\usepackage{dcolumn}
\usepackage{bm}
\usepackage{siunitx}

\newcommand{\microns}{\, \si{\micro \meter}}

\begin{document}


\title{Experimental investigation of the thermal emission cross-section of nano-resonators using hierarchical Poisson-disk distributions}

\author{Denis Langevin}
\thanks{AID funding}
\affiliation{DOTA, ONERA, Universit\'e Paris-Saclay, F-91123 Palaiseau - France}

\author{Clément Verlhac}
\affiliation{DOTA, ONERA, Universit\'e Paris-Saclay, F-91123 Palaiseau - France}

\author{Julien Jaeck}
\affiliation{DOTA, ONERA, Universit\'e Paris-Saclay, F-91123 Palaiseau - France}

\author{Loubnan Abou-Hamdan}
\affiliation{DOTA, ONERA, Universit\'e Paris-Saclay, F-91123 Palaiseau - France}
\altaffiliation{Institut Langevin, ESPCI Paris, PSL University, CNRS, 1 rue Jussieu, F-75005 Paris, France}

\author{Nathalie Bardou}
\affiliation{Center for Nanosciences and Nanotechnology (C2N) - CNRS, Université Paris-Saclay, 10 Boulevard
Thomas Gobert, 91120 Palaiseau, France}

\author{Christophe Dupuis}
\affiliation{Center for Nanosciences and Nanotechnology (C2N) - CNRS, Université Paris-Saclay, 10 Boulevard
Thomas Gobert, 91120 Palaiseau, France}

\author{Yannick De Wilde}
\affiliation{Institut Langevin, ESPCI Paris, PSL University, CNRS, 1 rue Jussieu, F-75005 Paris, France}

\author{Riad Ha\"idar}
\affiliation{DOTA, ONERA, Universit\'e Paris-Saclay, F-91123 Palaiseau - France}

\author{Patrick Bouchon}
\email {patrick.bouchon@onera.fr}
\affiliation{DOTA, ONERA, Universit\'e Paris-Saclay, F-91123 Palaiseau - France}

\date{\today}

\begin{abstract}
Effective cross-sections of nano-objects are fundamental properties that determine their ability to interact with light. However, measuring them for individual resonators directly and quantitatively remains challenging, particularly because of the very low signals involved. Here, we experimentally measure the thermal emission cross-section of metal-insulator-metal nano-resonators using a stealthy hyperuniform distribution based on a hierarchical Poisson-disk algorithm. In such distributions, there are no long-range interactions between antennas, and we show that the light emitted by the metasurface behaves as the sum of cross-sections of independent nanoantennas, enabling direct retrieval of the single resonator contribution. 
The emission cross-section at resonance is found to be of the order of $\mathbf{\lambda_0^2/3}$, a value that is nearly three times larger than the theroretical maximal absorption cross-section of a single particle but remains smaller than the maximal extinction cross-section.
This measurement technique can be generalized to any single resonator cross-section, and we also apply it here to the extinction cross-section.
\end{abstract}

\maketitle

Optical nanoantennas, often inspired by radiofrequency designs of classical antennas \cite{munk2003finite}, have been used to tame electromagnetic wave properties at optical wavelengths far below the diffraction limit \cite{brongersma2008engineering,novotny2011antennas,biagioni2012nanoantennas,krasnok2013optical,koenderink2017single}. 
The interactions of single nanoantennas with light can be of great interest to a multitude of light applications \cite{neubrech2008resonant,brongersma2015plasmon,mann2016quantifying,taylor2017single,zhu2019near,gillibert2020nanospectroscopy,sakat2020enhancing,rajabali2022ultrastrongly}, but they can also be used as building blocks (meta-atoms) for larger metasurfaces \cite{meinzer2014plasmonic,lalanne2017metalenses}. 
Among the fundamental properties of single objects are their effective absorption and scattering cross-sections, the sum of which is the extinction cross-section \cite{bohren2008absorption,staude2013tailoring}. These quantities describe the single object's ability to interact with light at a given frequency but are often complex to quantify experimentally \cite{Husnik2012quantitative,Crut2014optical}.

In practice, the direct measurement of the absorption cross-section has so far remained rather elusive \cite{baffou2012thermal}. As a result, the absorption cross-section is typically obtained by taking the difference between the extinction and scattering cross-sections.  
The emission cross-section can also be introduced based on the local polarized emissivity density that is linked to absorptivity by Kirchhoff's law \cite{Greffet2018light}. Direct measurement of infrared thermal emission of single objects has been previously demonstrated but the radiated signal is very low, particularly in comparison with the emitting background \cite{de2006thermal,li2018near}. 
Determining absorption and emission cross-sections of single objects is of the utmost importance for applications, such as thermoplasmonics \cite{jauffred2019plasmonic}, energy conversion, \cite{cuevas2018radiative,li2018nanophotonic} and meta-emitters with tailored emission (polarization, spectral) \cite{makhsiyan2015shaping,baranov2019nanophotonic,wojszvzyk2021incandescent,garcia2022tunable,nguyen2022circularly}.

In this letter, we measure, both directly and quantitatively, the thermal emission cross-section of nanoresonators arranged as a correlated disorder distribution.
The metasurface is built following a hierarchical Poisson disk algorithm with various densities that gives a stealthy hyperuniform distribution. For low densities, we show that the thermal emission is linearly dependent on the density of nanoantennas, thus behaving as an independent combination. In that case, we measure the sum of the individual antennas' thermal emission cross-sections. 
These measurements are shown to be in fair agreement with electromagnetic computations on a single nanoantenna.
Eventually, we demonstrate that our experimental protocol is easily adaptable to other cross-section measurements by applying it to the measurement of extinction cross-sections.

\begin{figure}[!ht]
\centering
\includegraphics[width=0.95\linewidth]{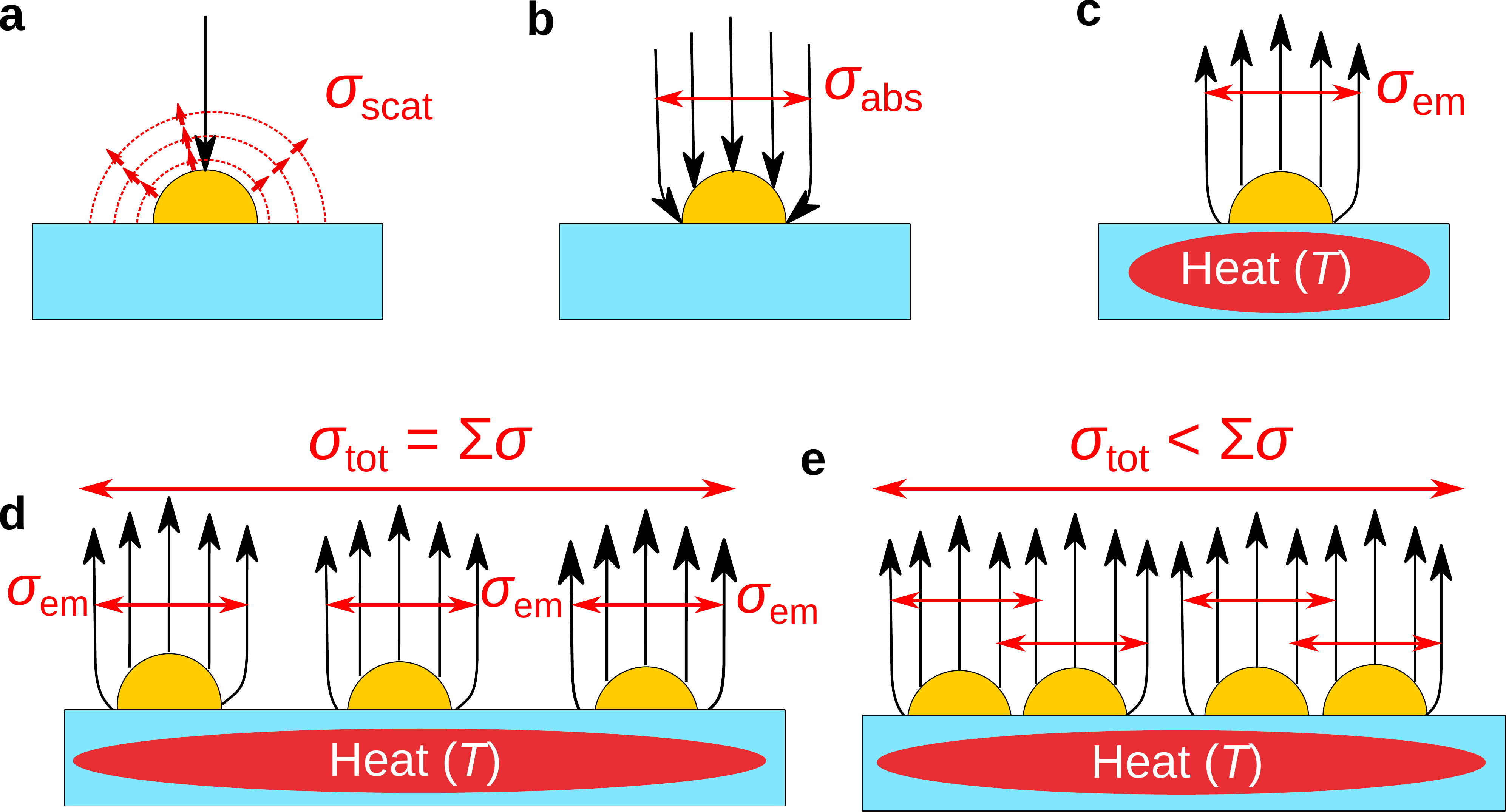}
\caption{ Illustration of the (a) scattering, (b) absorption and (c) emission cross-sections for a nano-object. (d) Non-overlapping cross-sections (e) Overlapping cross-sections.}
\label{Fig:00}
\end{figure}

The various electromagnetic cross-sections of a single nano-object on an opaque substrate are illustrated in  Fig.~\ref{Fig:00}(a-c). The scattering cross-section $\sigma_{scat}$ and the absorption cross-section $\sigma_{abs}$ are shown for the case of a normally incident monochromatic plane wave. Their sum gives the well-known extinction cross-section $\sigma_{ext}=\sigma_{abs}+\sigma_{scat}$.
At a temperature T, the nano-object emits light due to thermal fluctuations according to Planck's law. Thus, we introduce the emission cross-section $\sigma_{em}$ that describes its ability to emit light for a given wavelength $\lambda$, direction of emission $\vec{u}$ and polarization.
The local form of Kirchhof's law equates the absorption cross-section to the emission cross-section for similar conditions, i.e., $\sigma_{em}(\vec{u}, \lambda)=\sigma_{abs}(\vec{u}, \lambda)$\cite{Greffet2018light}.

An object's emission cross-section in a given direction $\vec{u}$ at a wavelength $\lambda$ is defined by:
\begin{equation}
    \sigma_{em}(\vec{u}, \lambda) = \int_V  \eta(\vec{u}, \vec{r}, \lambda)d^3\vec{r}
\end{equation}
where $\eta(\vec{u}, \vec{r}, \lambda)$ is the local spectral directional emissivity at position $\vec{r}$ in the object. This is equal to the absorptivity of the object and can be computed using the object's local permitivitty \cite{Greffet2018light}.

Figure~\ref{Fig:00}(d-e) extends this illustration to assemblies of nano-objects emitting light. In the first case, the nano-objects are sufficiently far apart so that no interactions may arise. The total emission cross-section $\sigma_{em}^{(tot)}$ is then given by the sum of the individual emission cross-sections $\sum\sigma_{em}^{(res)}$. In the second case (Fig.~\ref{Fig:00}(e)), the resonators are much closer to one another and can no longer be considered independent. Therefore, the individual cross-sections become spatially overlapping, giving rise to near-field interactions so that $\sigma_{em}^{(tot)}\neq\sum\sigma_{em}^{(res)}$. In most situations, the overlap of cross-sections leads to a loss of efficiency, \textit{i.e.,} $\sigma_{em}^{(tot)}<\sum\sigma_{em}^{(res)}$.
However, hybridized modes may appear and generate constructive interferences, leading to $\sigma_{em}^{(tot)}>\sum\sigma_{em}^{(res)}$.
In any case, the total far-field emission obviously remains below the fundamental limit set by the fact that the emissivity of any object cannot be larger than 1. 

Periodic arrays of nanoresonators are commonly employed to reach this theoretical limit, in which  complete absorption of incoming light is readily achieved by reaching a critical coupling condition \cite{Cui2014plasmonic}.
 
However, the array period, along with diffraction effects, play a major role in the overall behavior of the sample, concealing the contribution of the individual resonators \cite{sterl2021shaping}. The emission cross-section of a nanoantenna in an array of period $d$ at critical coupling can be written as $\sigma_{em}^{(tot)}=d^2$ at resonance. Consequently, the measurement of electromagnetic cross-sections can only be effected in the case of a single resonator or an aperiodic resonator distribution.

\begin{figure}[!ht]
\centering
\includegraphics[width=0.9\linewidth]{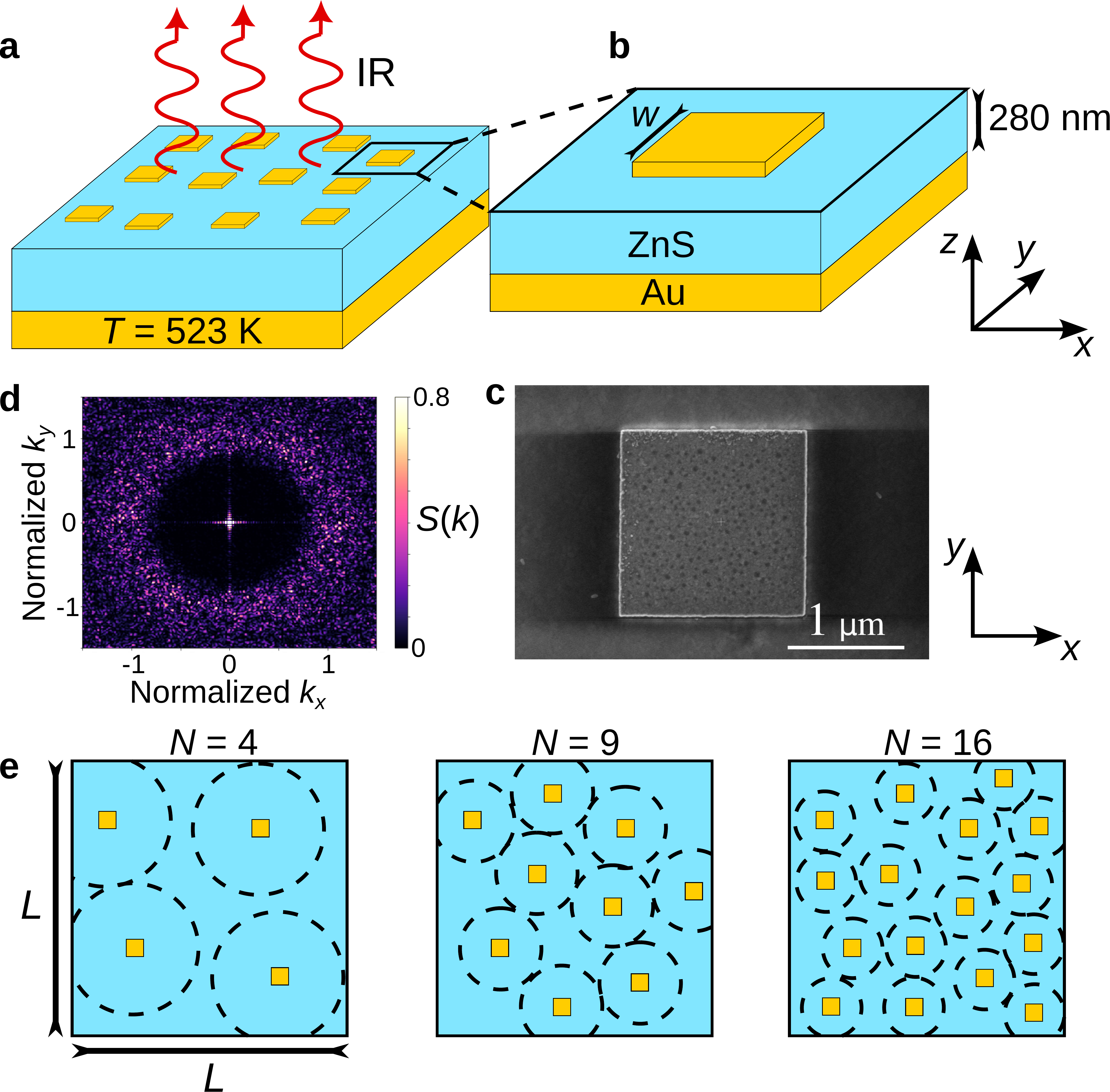}
\caption{(a) Scheme of the sample, with the overall distribution. (b) Scheme of the single resonator geometry. (c) Scanning electron microscope (SEM) top view image of a single resonator ($w=1.6 \microns$). (d) Structure factor of a distribution where $N=40000$ and $L = 1$ mm. (e) Examples of three consecutive states of the resonator distribution, after N steps. The dashed circles are exclusion zones where no additional antennas can be placed.}
\label{Fig.0}
\end{figure}

In what follows, the emission response of hyperuniform disordered metasurfaces using metal-insulator-metal (MIM) resonators as building blocks is investigated, as depicted in Fig.~\ref{Fig.0}. This type of antenna, commonly used in metasurface design, acts as a Fabry-Perot nano-cavity \cite{Cui2014plasmonic,Ogawa2018metal}. Here, the MIM resonators consist of square gold antennas placed on top of a zinc sulfide (ZnS) layer with a metallic backplate. The thickness of the dielectric layer is 280 nm, and the antenna width $w$ is chosen so that the resonance wavelength is in the long-wave infrared spectral range (8-12$\microns$). The metallic layer thickness is chosen to be significantly larger than the skin-depth of the metal. A scheme of this resonator structure is shown in Fig.~\ref{Fig.0}(a-c), together with an SEM image of a single resonator.

The stealthy hyperuniform disordered distribution is obtained following a hierarchical Poisson-disk algorithm \cite{McCool1992hierarchical}. This algorithm resembles Random Sequential Adsorption (RSA) \cite{widom1966random}, in that it builds the distribution gradually by randomly picking new positions one by one, only adding a new position to the distribution if it satisfies a pre-defined separation distance with respect to all previously chosen positions. The hierarchical Poisson-disk algorithm introduces a large exclusion distance between new positions, that is gradually decreased as the distribution density $D$ increases. 

This ensures that the algorithm can place a precisely chosen number of resonators $N$, rather than stopping when the exclusion zones overlap, and that the $n$ chosen positions first are also hyperuniform, for any $n<N$. 
Thus, once a design of given density $D$ is created, designs of any density $d<D$ can be retrieved by simply taking a subset of the densest distribution. This guarantees that the density variations between samples are only caused by adding or removing resonators from an existing distribution, without changing the positions of other resonators in the process.
This type of distribution minimizes the long-range correlations between resonators, thereby precluding the contribution of structural effects of the distribution on the signal \cite{torquato2003local, vynck2021light}.
A way to characterize hyperuniformity is through the structure factor of the resonator distribution that can be computed for any distribution $\{R\}_j$ of $N$ points as \cite{vynck2021light}:
$$
S(\Vec{k}) = \frac{1}{N} \sum_{j_1} \sum_{j_2} \exp^{-i\Vec{k}(\Vec{R_{j_1}}-\Vec{R_{j_2}})}.
$$
The structure factor for a hierarchical Poisson-disk distribution with $N=40000$ is plotted in Fig.~\ref{Fig.0}(d). The dark central area in this structure factor is a typical characteristic of a stealthy hyperuniform distribution \cite{Yu2021engineered}. 
The absence of high intensity peaks outside of the dark area indicates that there are no long-range correlations between positions.
An example of the state of the distribution given by the hierarchical Poisson-disk algorithm at different steps is shown in Fig.~\ref{Fig.0}(e). 
The overall resonator density $D$ of each sample is determined by the number of resonators per sample area: $D= N/L^2$. We can identify samples by their resonator density $D$, or by the period $L_{eq} = 1/\sqrt{D}$ of its equivalent periodic array of the same density.

\begin{figure}[!ht]
\centering
\includegraphics[width=0.8\linewidth]{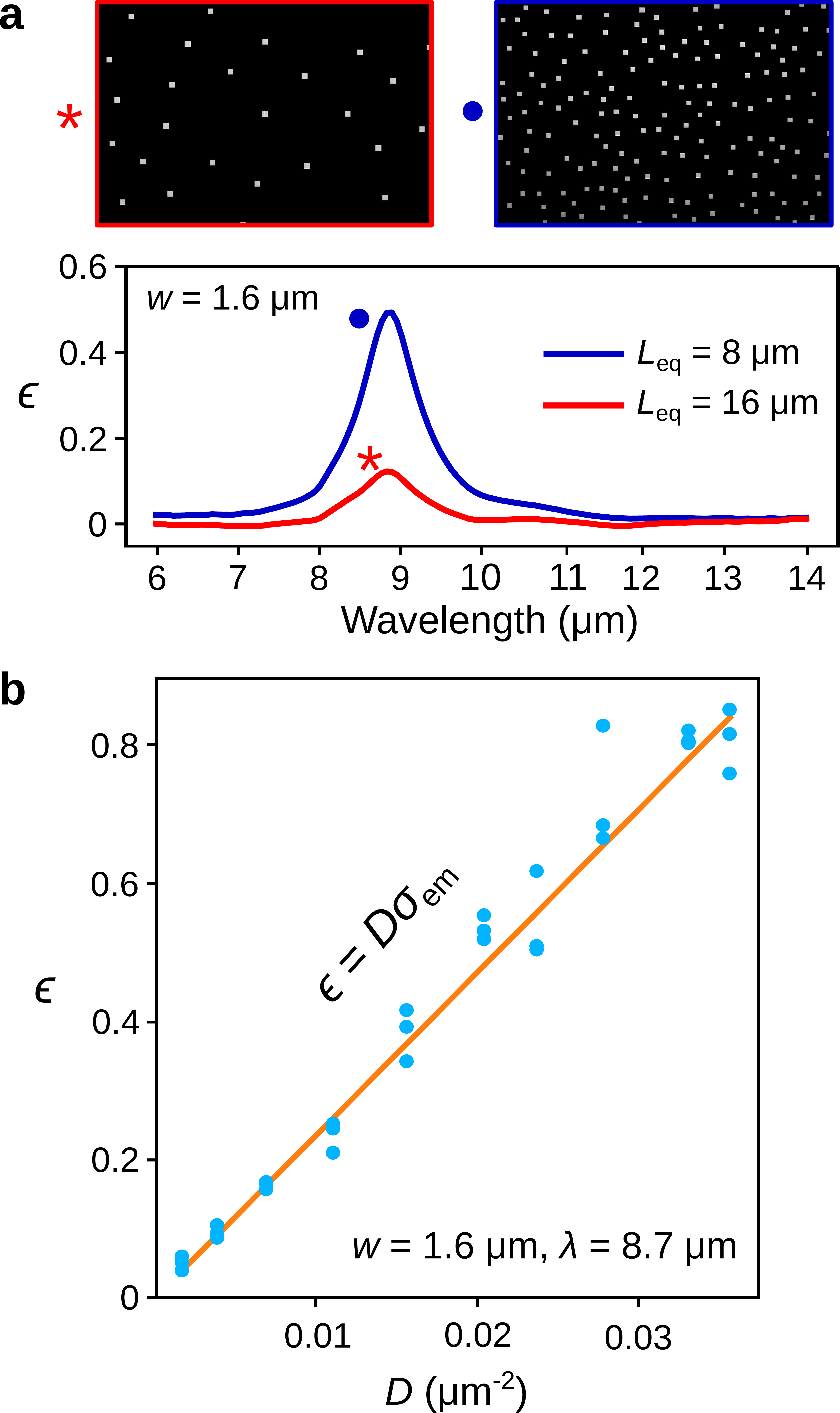}
\caption{(a) Measured emissivity $\epsilon$ spectra of the hyperuniform disordered metasurface with resonator density $D=1/64\microns^{-2}$, or equivalent period $L_{eq}=8\microns$ (blue curve) and $D=1/256\microns^{-2}$, or  $L_{eq}=16\microns$ (red curve). The spectra were measured at $T=523 \,\mathrm{K}$. The SEM images of the two metasurfaces are shown above at the same magnification. (b) Emissivity as a function of the density of the metasurface at resonance wavelength $\lambda _r =8.7 \microns$ for several measurements (blue dots). The linear regression slope (orange line) gives the emission cross-section $\sigma_{em}$ at $\lambda _r=8.7 \microns$.}
\label{Fig.1}
\end{figure}

The infrared thermal emission of disordered MIM metasurfaces with various densities, heated at $T=523 \,\mathrm{K}$ with a Linkam THMS600 heating stage, is collected around normal incidence by a parabolic mirror. The heated metasurface acts as an external source for a Fourier transform infrared spectrometer (FTIR) that measures its emission spectrum. 

The studied metasurfaces consist of square-patch MIM antennas of width $w=1.6 \microns$ distributed according to the hierarchical Poisson-disk distribution at various densities. 
The emissivity spectra of metasurfaces with resonator density $D=1/64\microns^{-2}$ and $D=1/256\microns^{-2}$ are shown in Fig.~\ref{Fig.1}(a), with scanning electron microscope (SEM) images illustrating the distribution. Both metasurfaces exhibit a resonance at the same wavelength $\lambda _r=8.7 \microns$, and as expected, the measured emissivity for the denser metasurface is higher. 

At each wavelength $\lambda$, the emission cross-section $\sigma_{em}(\lambda)$ can be extracted from the slope of the linear regression of the emissivity as a function of density. Indeed, as illustrated in Fig.~\ref{Fig:00}(d), for independant nano-emitters, the emissivity is given by $\epsilon = \sigma_{em} D$.
This behavior is illustrated at the resonance wavelength in Fig.~\ref{Fig.1}(b). 
The measured value gives $\sigma_{em}=24\microns^{2}$. This means that $\sigma_{em}\sim \frac{\lambda_r^2}{3}$ for the resonant wavelength of 8.7$\microns$. This order of magnitude is to be expected for a nanoresonator on a reflective backplate \cite{Husnik2012quantitative,striebel2017absorption}, but this value is higher than the maximum theoretical value for a single resonant dipolar particle, \textit{i.e.,} $\sigma _{\mathrm{abs}}^{\mathrm{dipole}} =\frac{3 \lambda ^2}{8 \pi}$ \cite{streed2012absorption,tretyakov2014maximizing}.

\begin{figure}[!ht]
\centering
\includegraphics[width=0.95\linewidth]{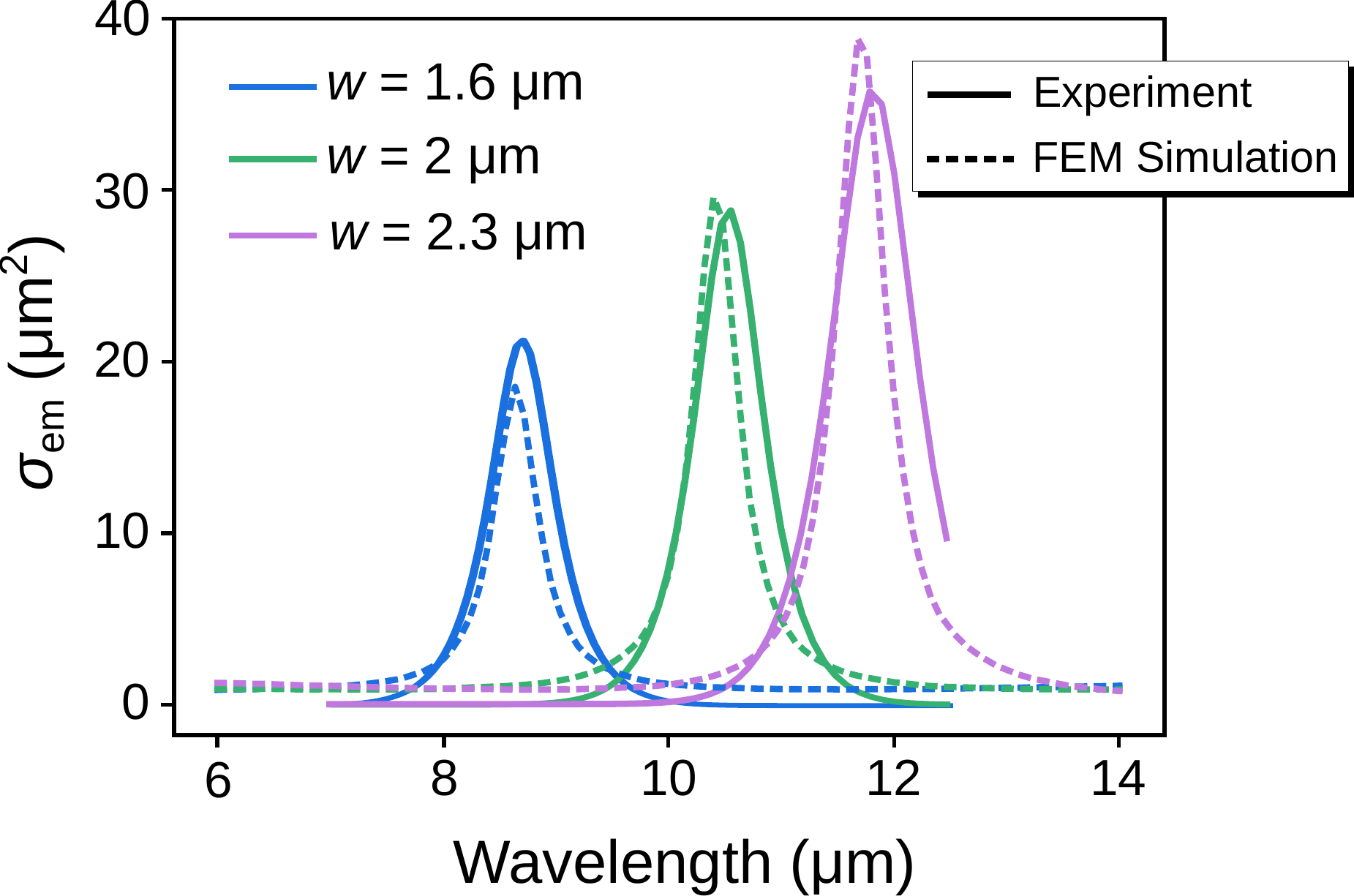}
\caption{Comparison of measured and simulated emission cross-section spectra. The measurements correspond to a resonator distributions of density $L_{eq}=16\microns$, while the electromagnetic simulations are for single resonators using perfectly matched layers.}
\label{Fig.3}
\end{figure}

Following the same process at each wavelength, we obtain the emission cross-section spectra, which are are shown on Fig.~\ref{Fig.3} for various resonator geometries. This shows that both the resonance wavelength and the maximum cross-section increase with the resonator width $w$, as expected \cite{li2018near}.

To further validate these experimental measurements, we performed finite element method (FEM) electromagnetic simulations (COMSOL Multiphysics \cite{multiphysics1998introduction}) of single resonators using perfectly matched layers. The emission cross-section of an individual square MIM resonator was retrieved and compared to the experimental measurements in Fig.~\ref{Fig.3}.
The FEM simulations show very good agreement with the experimental measurements, both in resonance wavelength and magnitude.

\begin{figure}[!ht]
\centering
\includegraphics[width=0.8\linewidth]{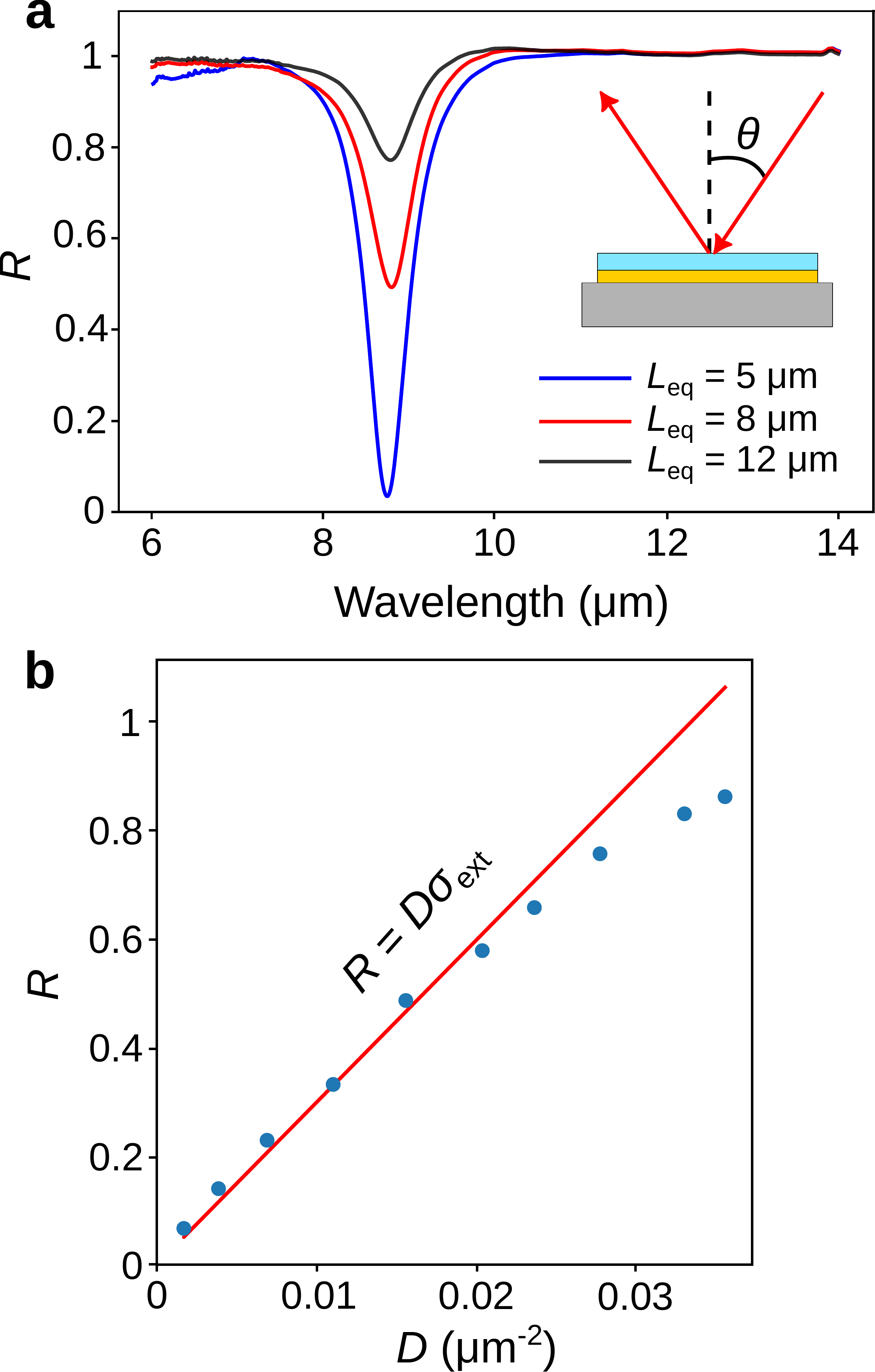}
\caption{(a) Measured reflectivity spectra for a resonator density $D=1/25\microns^{-2}$, or equivalent period $L_{eq}=5\microns$ (blue),  $D=1/64\microns^{-2}$, or  $L_{eq}=8\microns$ (red) and $D=1/144\microns^{-2}$, or  $L_{eq}=12\microns$ (green). The measurements are shown for an incidence angle $\theta = 13^{\circ}$ (see inset). (b) Linear regression of measured extinction against density to compute the extinction cross-section.}
\label{Fig.4}
\end{figure}

We underscore here that the measurement protocol that we have presented, using stealthy hyperuniform samples of varying antenna densities, is extremely robust and can be generalized to any kind of cross-section.
Concordantly, we have also used it to measure extinction cross-sections. In this case, reflectivity measurements were performed at an angle of incidence $\theta = 13 ^\circ$.
Figure~\ref{Fig.4}(a) shows the measured reflectivity spectra for three different sample densities, showing a dip in reflectivity at the resonance, representing the extinction by the individual nanoresonators that increases with increasing density.
The extinction values measured as a function of density (Fig.~\ref{Fig.4}(b)) follow a clear linear behavior at low densities. This leads to $\sigma_{ext} = 30\microns^2$, a value close to $\frac{2 \lambda ^2}{5}$. Contrary to the emission cross-section, this value is lower than the maximum theoretical value for a dipolar particle ($\sigma _{\mathrm{ext}}^{\mathrm{dipole}}=\frac{3 \lambda ^2}{2 \pi}$).  
Saturation can be observed at higher densities, where the measured extinction is no longer proportional to the density of resonators. This behavior originates from the overlap between resonator cross-sections, decreasing the efficiency of each individual resonator.
In the current study, this was only observed for the extinction measurements, as the resulting cross-section is larger than that obtained from emissivity measurements. The highest sample densities we characterized were sufficient to result in an extinction cross-section overlap but were still too low to give rise to an emission cross-section overlap.

In summary, we have presented a protocol to quantitatively measure single resonator cross-sections using large distribution far-field measurements. We applied this protocol to emission cross-sections of MIM resonator distributions, for which we obtained $\sigma_{em}=24\microns^2$.
These results were found to be in fair agreement with FEM simulations of single resonators, confirming that our measurement protocol gives direct access to the single resonator emission cross-section.
Finally, we extended our measurement protocol to extinction cross-sections. In this case, a saturation behavior was observed at high densities, which stresses the importance of performing measurements at various sample densities in order to properly isolate the linear regime, in which the presented protocol is applicable.

\bibliography{biblio_sigmaeff}

\end{document}



\title{Supplemental Material : Experimental investigation of the thermal emission cross-section of nano-resonators using hierarchical Poisson-disk distributions}

\begin{abstract}
We introduce the experimental bench and protocol for emissivity measurements in a first part. Then, we provide information on the fabrication of samples and on the numerical simulation of the EM response of a periodic resonator array and of a single resonator. Finally, we compare structure factors of various distributions.   
\end{abstract}

\author{Denis Langevin}
\thanks{AID funding}
\affiliation{DOTA, ONERA, Universit\'e Paris-Saclay, F-91123 Palaiseau - France}

\author{Clément Verlhac}
\affiliation{DOTA, ONERA, Universit\'e Paris-Saclay, F-91123 Palaiseau - France}

\author{Julien Jaeck}
\affiliation{DOTA, ONERA, Universit\'e Paris-Saclay, F-91123 Palaiseau - France}

\author{Loubnan Abou-Hamdan}
\affiliation{DOTA, ONERA, Universit\'e Paris-Saclay, F-91123 Palaiseau - France}
\altaffiliation{Institut Langevin, ESPCI Paris, PSL University, CNRS, 1 rue Jussieu, F-75005 Paris, France}

\author{Nathalie Bardou}
\affiliation{Center for Nanosciences and Nanotechnology (C2N) - CNRS, Université Paris-Saclay, 10 Boulevard
Thomas Gobert, 91120 Palaiseau, France}

\author{Christophe Dupuis}
\affiliation{Center for Nanosciences and Nanotechnology (C2N) - CNRS, Université Paris-Saclay, 10 Boulevard
Thomas Gobert, 91120 Palaiseau, France}

\author{Yannick De Wilde}
\affiliation{Institut Langevin, ESPCI Paris, PSL University, CNRS, 1 rue Jussieu, F-75005 Paris, France}

\author{Riad Ha\"idar}
\affiliation{DOTA, ONERA, Universit\'e Paris-Saclay, F-91123 Palaiseau - France}

\author{Patrick Bouchon}
\email {patrick.bouchon@onera.fr}
\affiliation{DOTA, ONERA, Universit\'e Paris-Saclay, F-91123 Palaiseau - France}

\date{\today}

\maketitle
\newpage

\section{Experimental setup for emissivity measurement}

\begin{figure}[ht]
\centering
\includegraphics[width=0.75\linewidth]{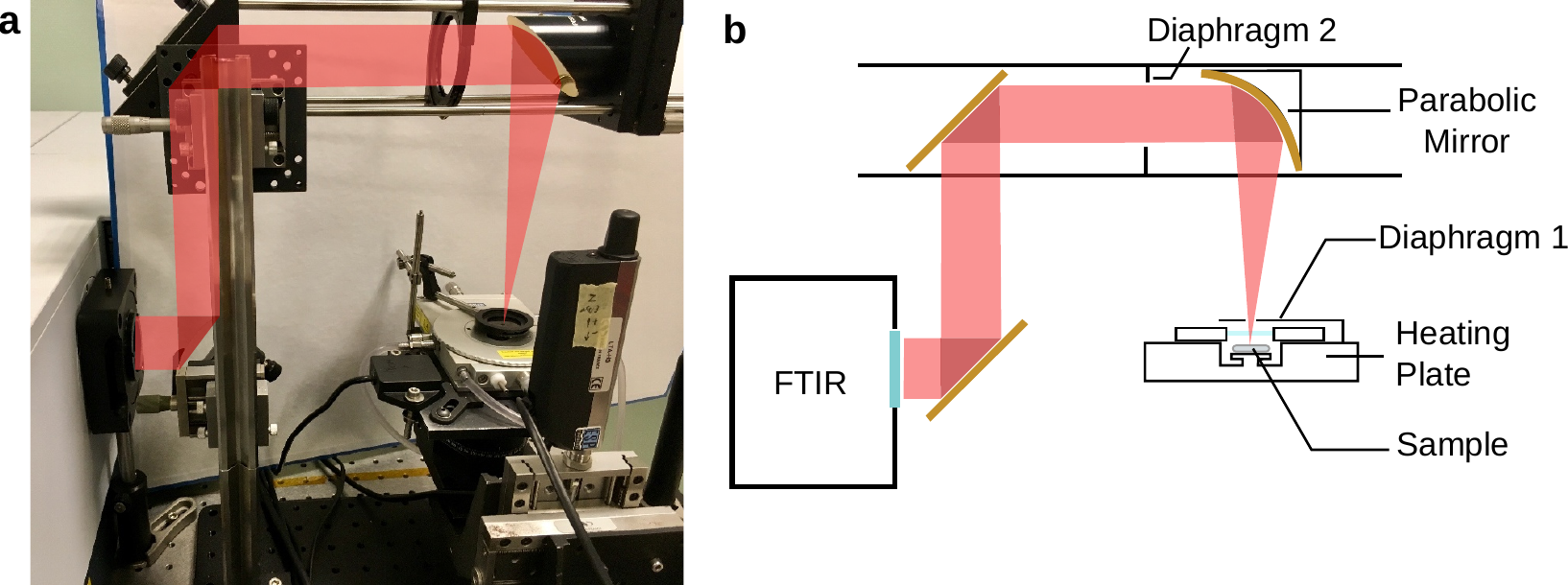}
\caption{(a) Picture of the thermal emission measurement setup, with the beam path represented in red that goes from the heating plate to the entrance of the Fourier transform infrared spectrometer (FTIR). (b) Scheme of the measurement setup, with its different parts.}
\label{Fig:Montage_thermique}
\end{figure}

The experimental protocol for the emissivity measurement is as follows: the sample is placed on a Linkam THMS600 heating stage (see  Fig.~\ref{Fig:Montage_thermique}) that controls the temperature, and the thermal emission signal from the heated sample $S_\mathrm{sample}$ is collected around normal incidence by a parabolic mirror, which directs it to a Bruker 70v Fourier transform interferometer (FTIR). In the FTIR, the signal goes through a Michelson interferometer and is detected by a HgCdTe detector with a cut-off wavelength at 16 µm. Thus, the sample plays the role of an external source to the FTIR, and to determine its emissivity, the background signal and a reference thermal source must also be measured in the same conditions.    
The thermal background $S_\mathrm{bg}$ is measured by replacing the sample by a mirror (200 nm gold-coated substrate) on the heating plate.
Finally, the thermal emission signal $S_\mathrm{ref}$ of a reference material (carbon black) of known emissivity $\epsilon _{\mathrm{CB}}$ is measured likewise, in order to compute the emissivity spectrum of the sample.
This emissivity $\epsilon(\lambda, \theta=0^\circ)$ can be written:
\begin{equation}
\epsilon = (\epsilon _{\mathrm{CB}} - \epsilon _{\mathrm{Au}} ) \frac{S_\mathrm{sample} - S_\mathrm{bg}}{S_\mathrm{ref}-S_\mathrm{bg}} + \epsilon _{\mathrm{Au}}.
\end{equation}

\begin{figure}[ht]
\centering
\includegraphics[width=0.75\linewidth]{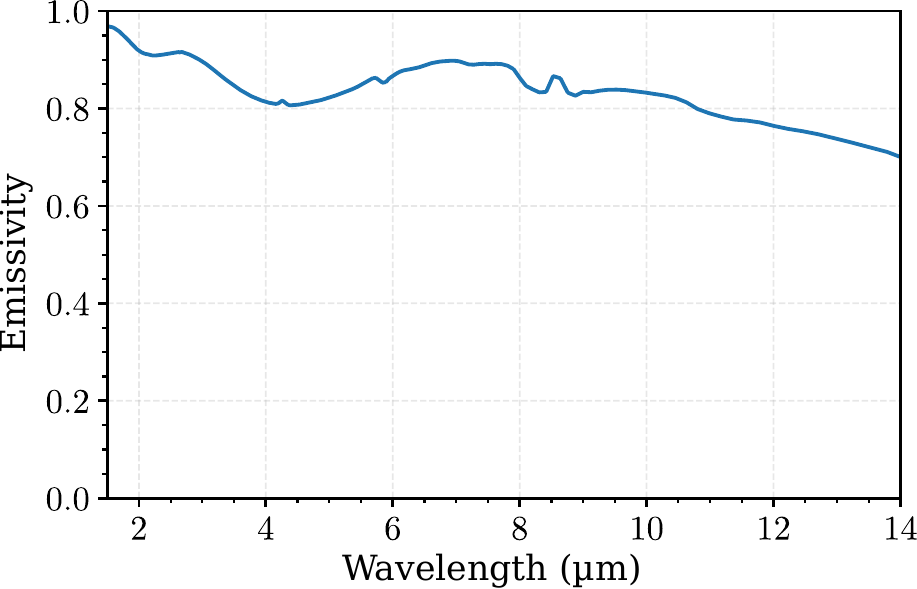}
\caption{Emissivity spectrum of carbon black ink.}
\label{fig:spectreCB}
\end{figure}

The reference material used is a carbon black ink containing 5\% of carbon black (JR-700HV - Metalon)  that is deposited on a Si wafer. The ink is then heated on a hot plate at $T=530 \,K$ for 5 minutes. It exhibits high optical absorption in the infrared \cite{chang1990determination,mizuno2009black}, with an emissivity close to 0.85 at the wavelengths considered here, as shown in Fig.~\ref{fig:spectreCB}.

\begin{figure}[ht]
\centering
\includegraphics[width=0.75\linewidth]{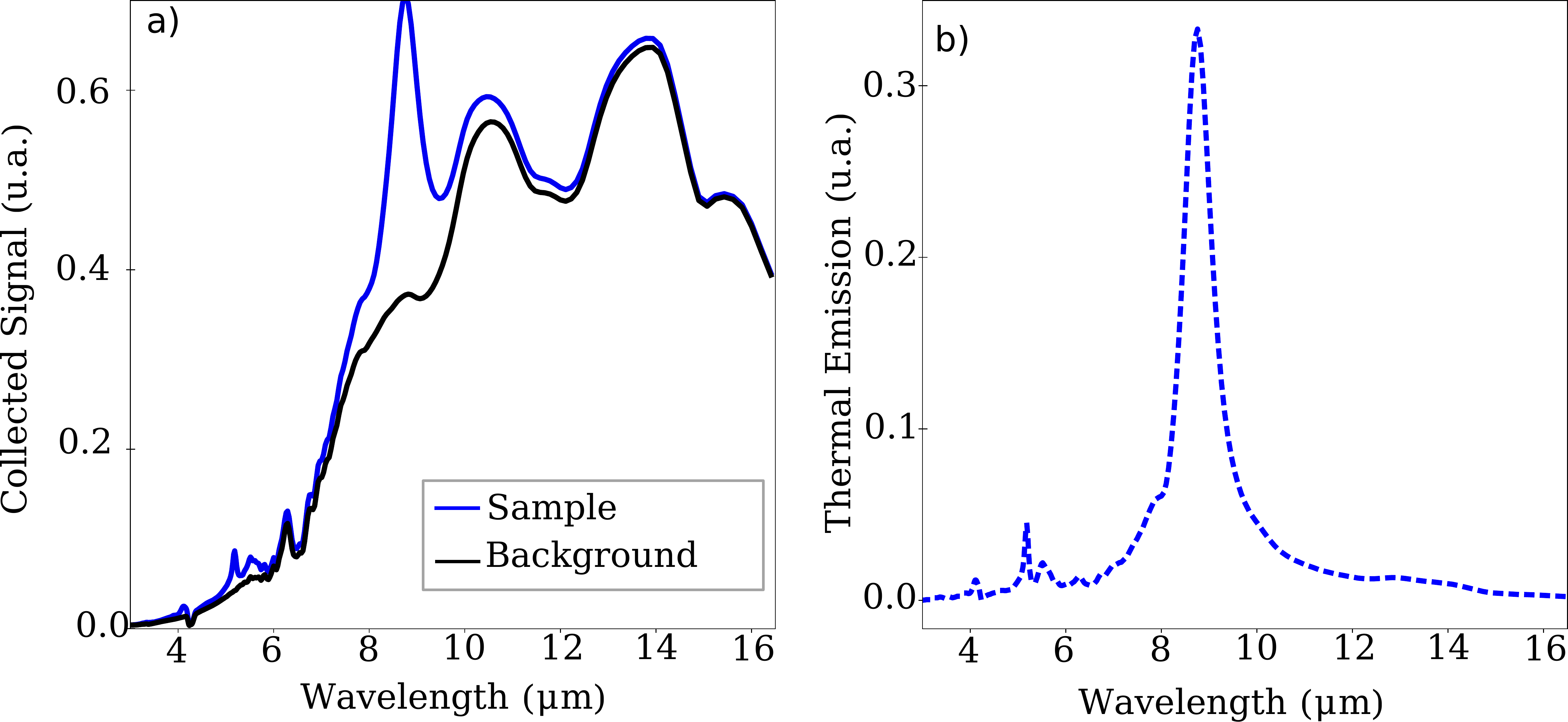}
\caption{Thermal emission measurements for resonators of width $w=1.6\microns$ and sample density $L_\mathrm{eq}=6\microns$, at 250°C. (a) Collected thermal emission signal. (b) Measured thermal emission spectrum computed as $S_\mathrm{sample} - S_\mathrm{bg}$.}
\label{Fig:Mesure_thermique}
\end{figure}

The thermal background can be higher than the signal we want to measure, as shown in the Fig.~\ref{Fig:Mesure_thermique}(a). This is due both to the overall ambient temperature emission, especially inside the interferometer, and to the heating plate emission.
This is however not a problem because this background is stable on the time scale of the measurement, and can therefore be measured and subtracted from the sample signal.
Figure~\ref{Fig:Mesure_thermique}(b) shows that at a heating temperature of 250°C, it is possible to measure the thermal emission spectrum $S_\mathrm{sample} - S_\mathrm{bg}$ with a good signal-to-noise ratio.

\begin{figure}[ht]
\centering
\includegraphics[width=0.75\linewidth]{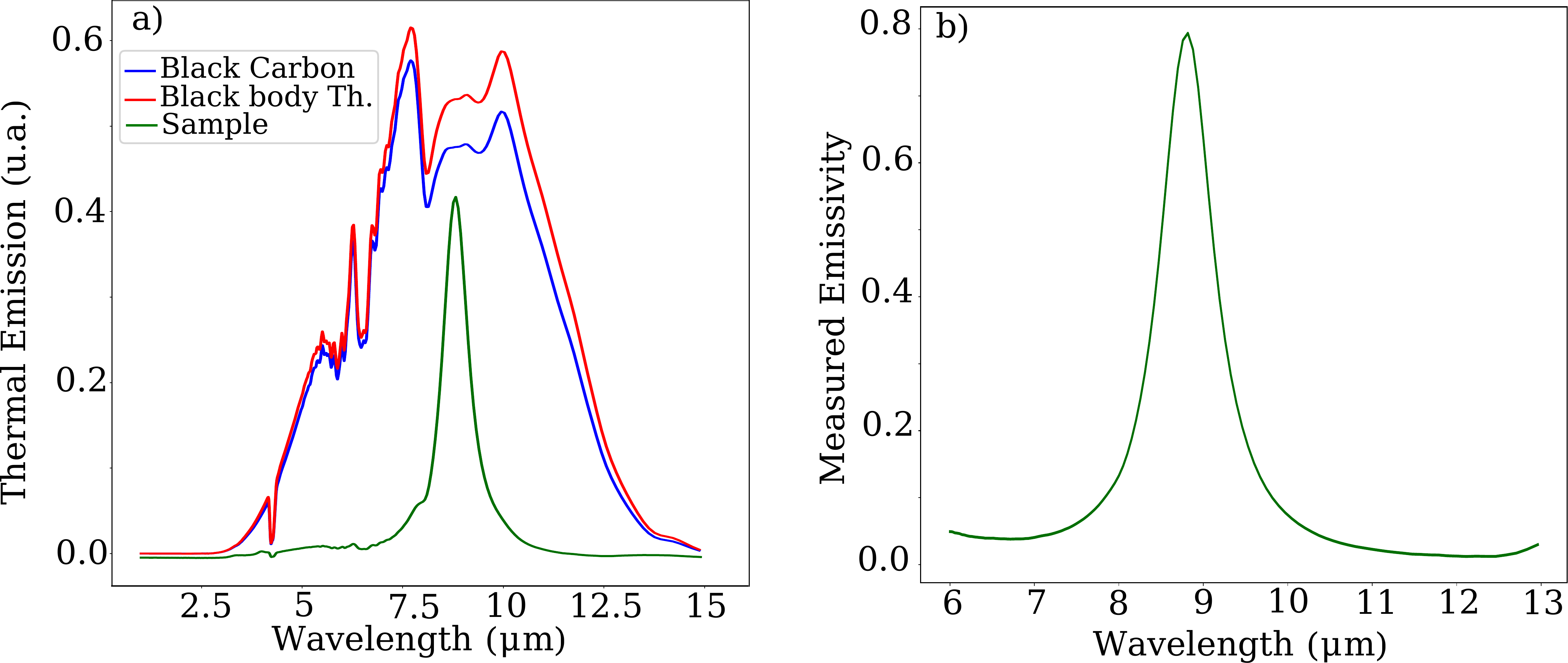}
\caption{(a) Measured thermal emission for a sample of hierarchical Poisson-disk distributed nanoantennas (with density $L_\mathrm{eq}=6\microns$, antenna length $w=1.6\microns$), as well as the carbon black reference signal. The maximum black body theoretical signal, computed from that of the carbon black, is also shown. (b) Emissivity spectrum retrieved from the measurements shown in (a).}
\label{fig:ex_emissivite}
\end{figure}

Figure~\ref{fig:ex_emissivite}(a) presents the measured thermal emission for the carbon black $S_\mathrm{ref} - S_\mathrm{bg}$, as well as the computed emission for a perfect blackbody, and compares them to the signal from a sample of density $L_\mathrm{eq}=6\microns$.
The computed emissivity spectrum $\epsilon(\lambda)$ is presented in Fig.~\ref{fig:ex_emissivite}(b).

\section{Electromagnetic simulation}

Two different simulation techniques are used in the main article. On the one hand, the COMSOL Multiphysics software \cite{multiphysics1998introduction} is used for single antenna cross-section simulations. On the other hand, the Reticolo software is used to compute the reflection spectra of periodic antenna arrays.

While the resonator widths $w$ vary between 1.6$\microns$ and 2$\microns$, the other dimensions are the same: ZnS substrate thickness of 280~nm, antenna thickness of 40~nm.

\subsection{Single resonator simulation with Comsol Multiphysics}
\begin{figure}[ht!]
\centering
\includegraphics[width=0.6\linewidth]{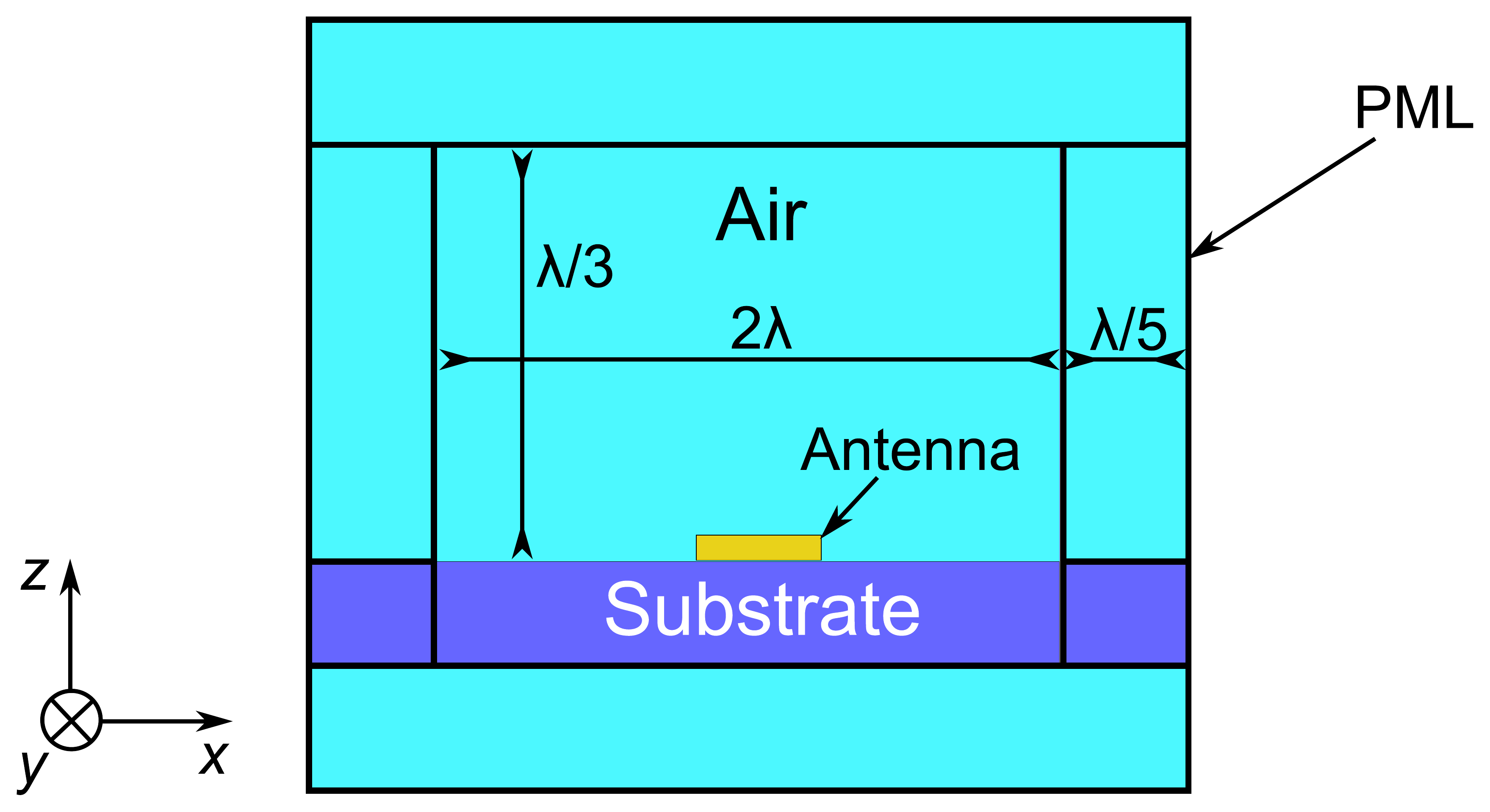}
\caption{Geometry used to simulate a single sub-$\lambda$ antenna using the commercial FEM solver of Maxwell's equations, Comsol Multiphysics. The geometry is bounded by perfectly matched layers (PMLs) of thickness $\lambda/5$ in all directions. }
\label{figA1}
\end{figure}

Finite element method (FEM) simulations, using the commercial software Comsol Multiphysics, were carried out to simulate cross-sections of individual subwavelength square-patch MIM antennas. To simulate a single antenna, \emph{perfectly matched layers} (PMLs)
were set for all boundaries of the geometry as illustrated in Fig.~\ref{figA1}. A plane wave port is incident from the top at an angle $\theta$ with respect to the normal to the substrate. The simulation is first run without the antenna, then run again with it so that the field is slightly modified by its presence, allowing to take into account the influence of the sub-$\lambda$ antenna on the large simulation region. The simulation span is typically taken to be $1.5\lambda$ or $2\lambda$, where $\lambda$ is the wavelength of the incident light, in order to reduce stray reflections off the boundaries and to ensure proper convergence of the simulations.  The dielectric functions of the materials making up the antenna were taken from the literature (ref.~\cite{olmon2012optical} for Au and ref.~\cite{querry1998optical} for ZnS).\par
The scattering cross-section of the antenna is calculated by integrating over the surface of the antenna as follows

\begin{equation}
    \sigma_{\textrm{scat}} = \frac{1}{I_{0}}\int_{S} \mathbf{n}. \mathbf{S}_{\textrm{scat}}dS,
\end{equation}

\noindent
where $\mathbf{n}$ is the normal vector pointing outward from the antenna, $\mathbf{S}_{\textrm{scat}}$ is the scattered
intensity (Poynting) vector, and $I_{0}$ is the intensity of the incident light.\par
The absorption cross-section, on the other hand, is obtained by integrating losses over the volume of the antenna as follows

\begin{equation}
    \sigma_{\textrm{abs}}  = \frac{1}{I_{0}}\int_{V}QdV,
\end{equation}

\noindent
where $Q$ is the power loss density in the antenna.\par
The extinction cross-section is then found by summing the scattering and absorption cross-sections:

\begin{equation}
    C_{\textrm{ext}}= C_{\textrm{abs}} +  C_{\textrm{scat}}.
\end{equation}

\subsection{Periodic resonator array simulation with Reticolo software}

Fourier Modal Method (FMM) simulations using the Reticolo software \cite{hugonin2021reticolo} were carried out to compute the reflectivity spectrum of periodic MIM arrays with a period $p=5\microns$. The geometry and materials parameters are the same as previously defined. 
The reflection spectra were computed with an incidence angle of 13$\degree$ and compared to the measured reflectivity spectrum with a fair agreement. There are no diffracted orders, so the incoming light is either reflected or absorbed, and the absorption is directly computed as $A=1-R$. A very high absorption, up to 96 \%, was obtained at the same resonance wavelength $\lambda=8.7 \microns$ as for disordered arrays. In that case, the equivalent emission cross-section would be $\sigma_{em}=24\microns^{2}$, but with a zero scattering cross-section. Besides, this value mostly depends on the period value. For instance, if $d=6\microns$, the simulations give $\sigma_{em}=27.7\microns^{2}$ and if $d=3 \microns$ then $\sigma_{em}=8.2\microns^{2}$.

\begin{figure}[ht]
    \centering
    \includegraphics[width=0.75\linewidth]{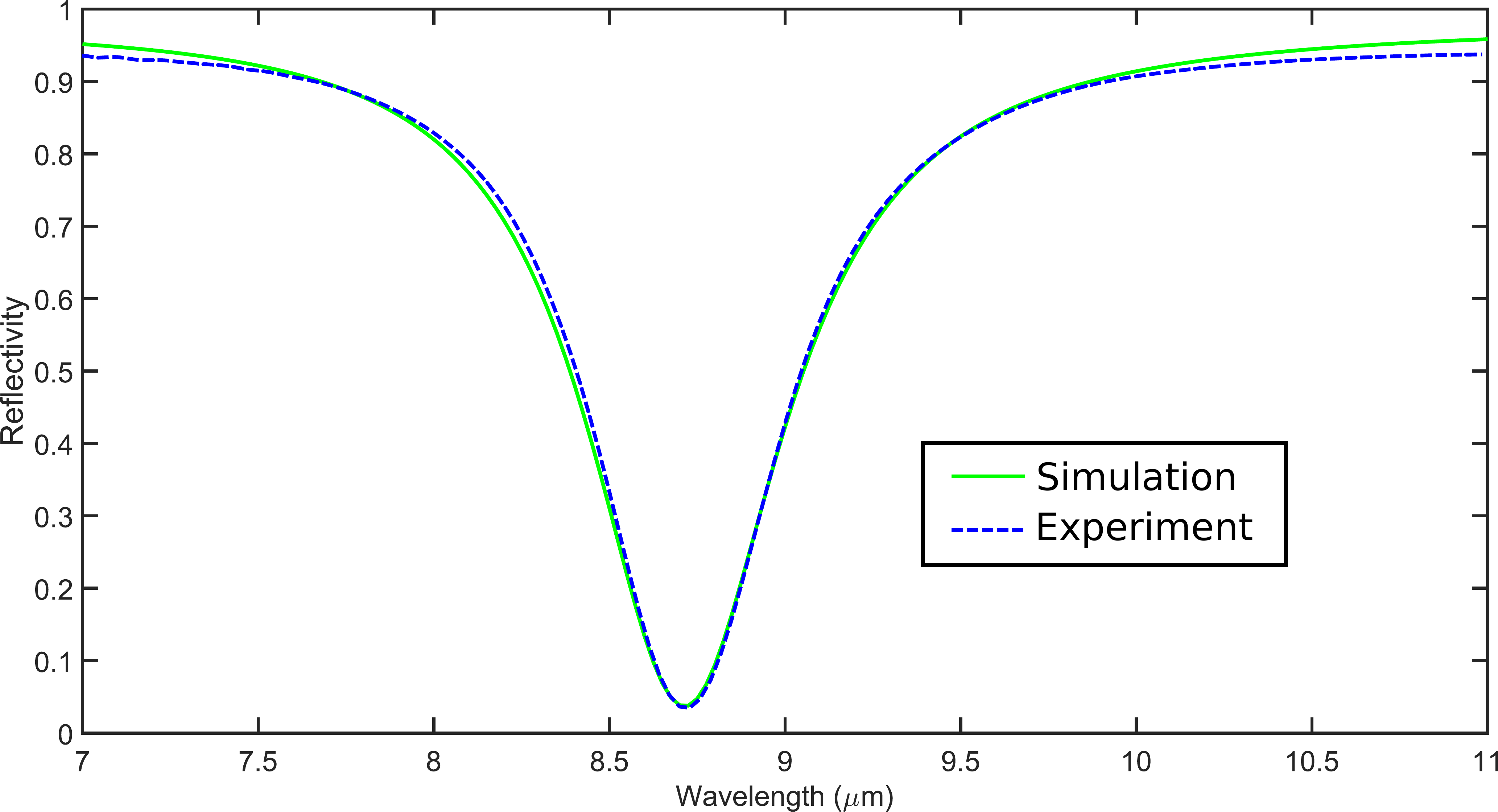}
    \caption{Simulated (green curve) and measured (dashed blue curve) reflectivity spectra for a periodic array of MIM resonators ($w=1.62\microns$, period $d=5 \microns$) at a 13$\degree$ incidence.}
    \label{fig:periodic}
\end{figure}

\section{Sample fabrication}
A gold layer of 200 nm is deposited on a silicon wafer with a 20 nm Ti adhesion layer. Then a subsequent 270 nm Zinc sulfide layer is deposited on top of it. 
The distributions of antennas are then obtained by an e-beam lithography with a subsequent lift-off. 
The samples were spin coated with PMMA A3 at 3000 rounds/minutes for 60 seconds and then baked at 170°C for 5 minutes. Then, e-beam lithography was carried out with a Vistec EBPG 5000 at 100 kV with a proximity effect correction that accounts for the electron scattering in the layers stack. The resist is developed in a solution of MIBK:IPA (1:3) for 70 seconds and rinsed in IPA for 10 seconds. Evaporation of 50 nm of gold with a 5 nm Cr adhesion layer was finally performed and the substrate was lifted in SVC-14 at 80°C.

\section{Structure factor of various distributions}

As stated in the main text, the structure factor of a distribution is a useful tool to determine its optical properties, especially the angular distribution of the light reflected off it.

\begin{figure}[ht]
    \centering
    \includegraphics[width=0.95\linewidth]{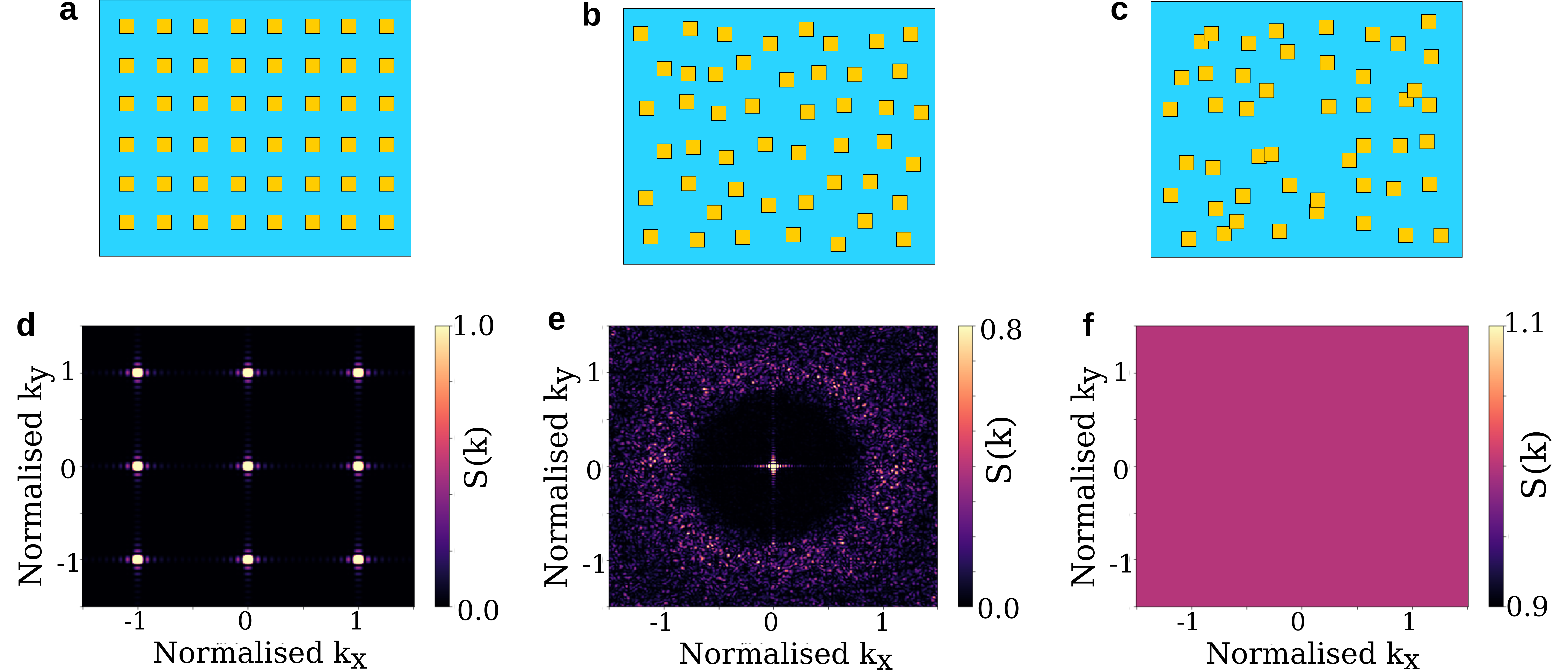}
    \caption{(top) Sample distributions and (bottom) structure factor of distributions of (a,d) periodic distribution, (b,e) hyperuniform distribution and (c,f) uncorrelated random distribution}
    \label{fig:FT_dist}
\end{figure}

Figure~\ref{fig:FT_dist} shows the differences between three distinct distribution types, by presenting both an example distribution and the typical structure factor of the distribution. 

 The first (a,d) is the periodic distribution, whose structure factor contains the typical very localized bright spots which are the signature of diffraction peaks in the reflected light.
 The second (b,e) is the hyperuniform distribution, which is the focus of the current paper, characterized by a central "dark" circle in its structure factor, a sign of the absence of long range correlations.
 The third (c,f) is the uncorrelated random distribution, where all positions are picked at random, and whose structure factor is uniform.
 
\bibliography{biblio_suppmat}